\documentclass[aps,prl,reprint,longbibliography,superscriptaddress]{revtex4-1}

\usepackage{amsmath,graphicx,color}
\usepackage{subfigure}
\usepackage{bm}
\usepackage[colorlinks=true,linkcolor=blue,citecolor=blue,urlcolor=blue]{hyperref}
\usepackage[all]{hypcap}
\usepackage{color}

		\newcommand{\rr}[0]{\boldsymbol{r}}
		\newcommand{\vv}[0]{\boldsymbol{v}}
		\newcommand{\nn}[0]{\boldsymbol{\nabla}}
		\newcommand{\JJ}[0]{\boldsymbol{J}}

		\newcommand{\Ds}[0]{D_\mathrm{s}}
		\newcommand{\Da}[0]{D_\mathrm{a}}
		\newcommand{\Rc}[0]{\rho_\mathrm{c}}
		\newcommand{\Tc}[0]{T_\mathrm{c}}
		\newcommand{\Te}[0]{T_\mathrm{eff}}
		\newcommand{\Ra}[0]{\rho_\mathrm{a}}
		\newcommand{\Ta}[0]{T_\mathrm{a}}
		\newcommand{\Nc}[0]{N_\mathrm{c}}
		\newcommand{\dd}[0]{\mathrm{d}}

\begin{document}

\title{Bose--Einstein Condensation in Scalar Active Matter with Diffusivity Edge}

\author{Ramin Golestanian}
\email{ramin.golestanian@ds.mpg.de}
\affiliation{Max Planck Institute for Dynamics and Self-Organization (MPIDS), 37077 G\"ottingen, Germany}
\affiliation{Rudolf Peierls Centre for Theoretical Physics, University of Oxford, Oxford OX1 3PU, United Kingdom}

\date{\today}

\begin{abstract}
Due to their remarkable properties, systems that exhibit self-organization of their components resulting from intrinsic microscopic activity have been extensively studied in the last two decades. In a generic class of active matter, the interactions between the active components are represented via an effective density-dependent diffusivity in a mean-field single-particle description. Here, a new class of scalar active matter is proposed by incorporating a diffusivity edge into the dynamics: when the local density of the system surpasses a critical threshold, the diffusivity vanishes. The effect of the diffusivity edge is studied under the influence of an external potential, which introduces the ability to control the behaviour of the system by changing an effective temperature, which is defined in terms of the single-particle diffusivity and mobility. At a critical effective temperature, a system that is trapped by a harmonic potential is found to undergo a condensation transition, which manifests formal similarities to Bose-Einstein condensation.
\end{abstract}


\maketitle
Dense active matter provides an enthralling paradigm as its emergent properties are determined by competition between opposing tendencies that originate from both equilibrium and non-equilibrium processes\cite{marchetti2013hydrodynamics,Bechinger:2016}. In particular, the proximity of the active agents -- through various non-equilibrium processes such as hydrodynamic interactions\cite{Elgeti:2015}, chemical signalling\cite{keller1971model,wadh04,taktikos+zaburdaev12,saha+golestanian14}, etc -- might lead to a collective enhancement of activity\cite{Sokolov:2007,Ishikawa:2011} and triggering of instabilities \cite{Simha:2002,Saintillan:2008,Golestanian:2012}, whereas short-ranged physical interactions such as those arising from stickiness and excluded volume effects (and possibly some non-equilibrium processes as well) inhibit the collective activity \cite{Cates:2015,Henkes:2011,Hagan:2013,Buttinoni:2013,soto+golestanian14}, ultimately leading to formation of globally ordered dense structures\cite{Blaschke:2016,Letitia:2018,Gompper:2018} or dynamic arrest\cite{Lu:2008,Cavagna:2009}. In the absence of long-range orientational ordering, which can be caused by alignment interactions between polar agents in sufficiently dense systems \cite{Toner:1998,gregoire2004onset}, the activity of individual particles can be described by effective enhanced diffusion coefficients beyond the time scale of rotational diffusion \cite{Howse:2007}. In this regime, the system can be generically described using dynamical equations for the density field at the mean field level. Here, we study such a description of scalar active matter with a generic density dependence in the diffusivity that incorporates a finite threshold: above a critical density the diffusivity vanishes. We demonstrate that the existence of this {\em diffusivity edge} leads to a dynamical phase transition that can be categorized as an analogue of Bose-Einstein condensation (BEC) despite the system being classical in nature. The present work builds on the recent surge in the development of generalized thermodynamic descriptions for non-equilibrium active matter\cite{Solon:2015,Grosberg:2015}.


We formulate a mean-field description of the dynamics of the colloidal system described by an effective single-particle density field $\rho({\rr},t)$ that satisfies a conservation law $\partial_t \rho+ \nn \cdot \JJ=0$
where the flux is defined as $\JJ=-D(\rho) \nn \rho+ \rho \vv$
in terms of an effective density-dependent diffusivity $D(\rho)$ and the drift velocity $\vv$. We assume that the drift originates from an external potential $U(\rr)$ and involves a density-dependent mobility $\mu(\rho)$, namely $\vv=\mu(\rho) (-\nn U)$. 
Using the single-particle diffusivity $\Ds=D(\rho \to 0)$ and the single-particle mobility $\mu_{\rm s}=\mu(\rho \to 0)$, we define an effective temperature $\Te$ using the fluctuation--dissipation theorem (FDT), namely, $k_{\rm B} \Te\equiv \Ds/\mu_{\rm s}$, and define $\beta \equiv 1/(k_{\rm B} \Te)$ for simplicity of notation. 

We incorporate a generic non-equilibrium character for the system by assuming a breakdown of FDT at finite densities, namely, $D(\rho)/\mu(\rho) \neq \Ds/\mu_{\rm s}$. We assume that this occurs due to density--dependent non-equilibrium effects, which could materialize as a result of collective inhibition as well as collective activation due to motility. Therefore, the diffusivity $D(\rho)$ and the mobility $\mu(\rho)$ will start from their single-particle values for dilute systems and can in general go up or down as the density is increased. To complete the formulation, we define a diffusivity edge at concentration $\Rc$ as follows: $D(\rho)/\mu(\rho)=0$ for $\rho \geq \rho_{\rm c}$. We note that in this work we ignore non-local effects that arise from hydrodynamic interactions when FDT is broken\cite{Golestanian:2002} as well as long-range non-equilibrium interactions \cite{Golestanian:2012}.

We then seek possible stationary states of the system as obtained by setting the net flux
\begin{equation}
\JJ/\Ds=- \big(D(\rho)/\Ds\big) \nn \rho-\big(\mu(\rho)/\mu_{\rm s}\big) \rho \nn \beta U, \label{eq:flux-2}
\end{equation}
to zero. This yields
$\frac{\dd \beta U}{\dd \rho}= - \frac{\mu_{\rm s} D(\rho)}{\Ds \mu(\rho) \rho}$, 
and consequently
\begin{equation}
\beta U (\rho)= - \int_{\rho_0}^{\rho} \frac{\dd \rho'}{\rho'} \;\frac{D(\rho')}{\Ds} \cdot \frac{\mu_{\rm s}}{\mu(\rho')}, \label{eq:U-1}
\end{equation}
where $\rho_0$ is defined as the density at which energy is at its lowest value of zero, namely, the ground state of the energy spectrum. The stationary distribution $\rho(U)$ can then be obtained by inverting equation (\ref{eq:U-1}). Henceforth, we are going to represent $\mu_{\rm s} D(\rho)/\big(\Ds \mu(\rho)\big)$ as $D(\rho)/\Ds$ to keep the presentation simple.

 \begin{figure*}[t]
 \includegraphics[width=0.27\linewidth]{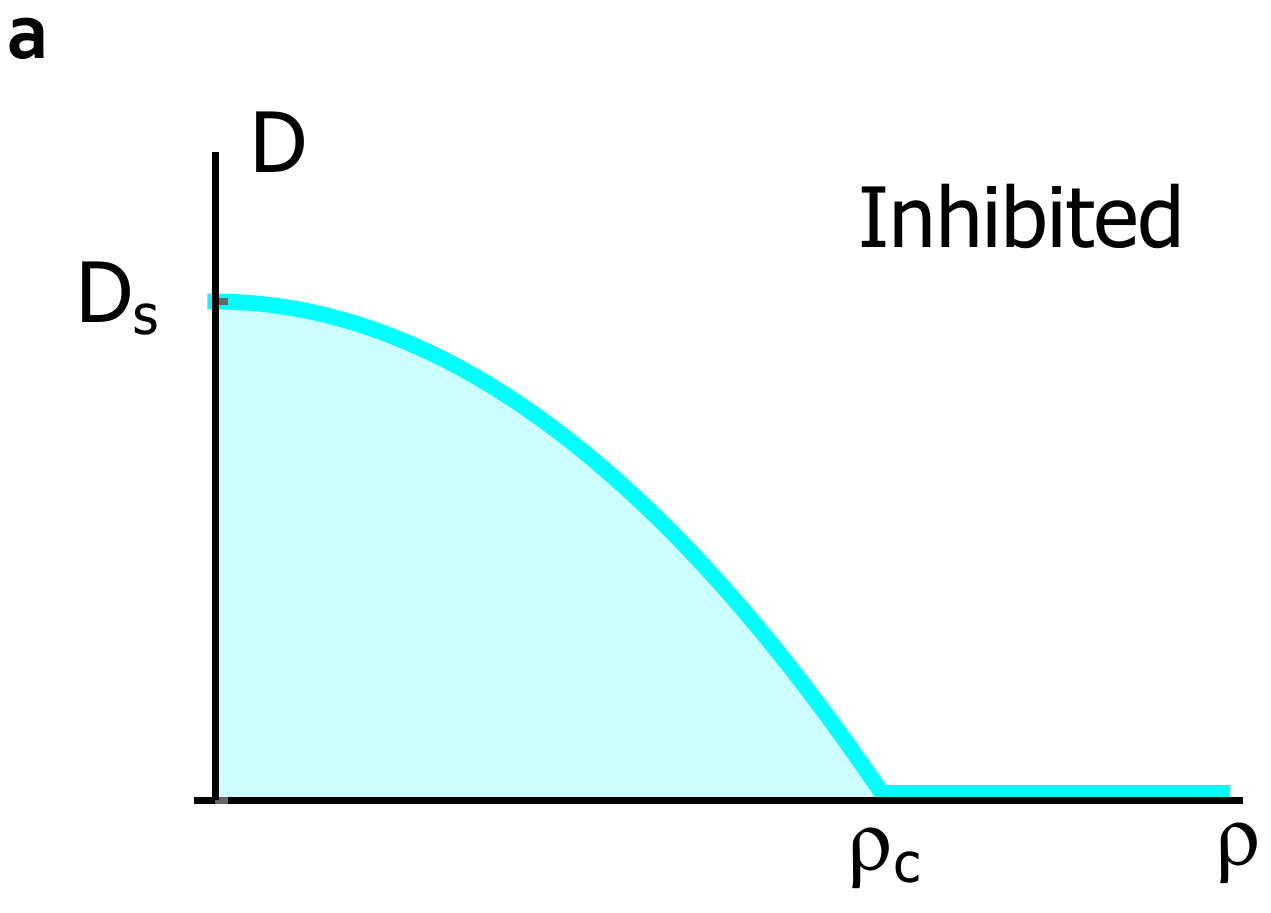} \hskip0.06\linewidth
  \includegraphics[width=0.24\linewidth]{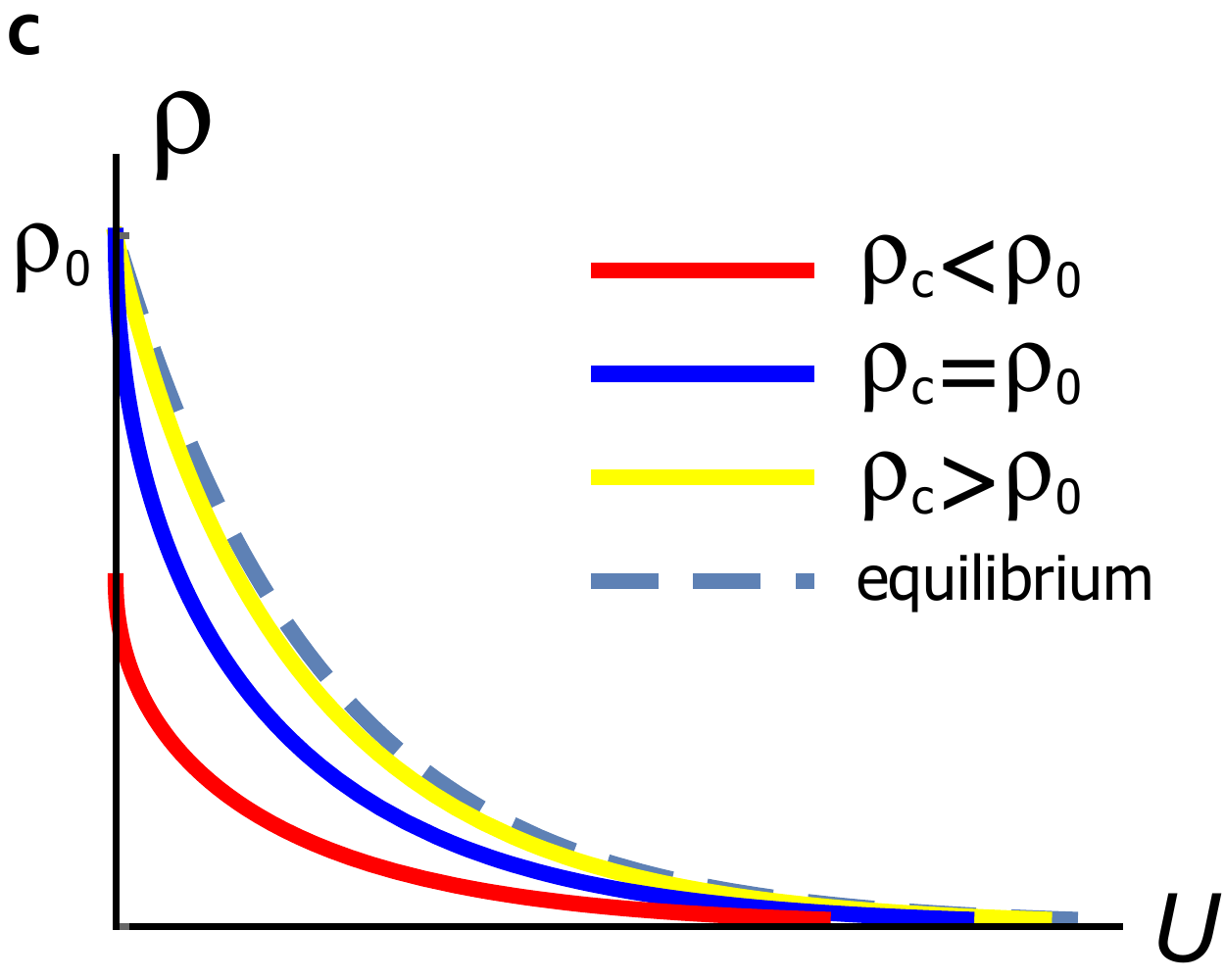} 
  \vskip5mm
    \includegraphics[width=0.22\linewidth]{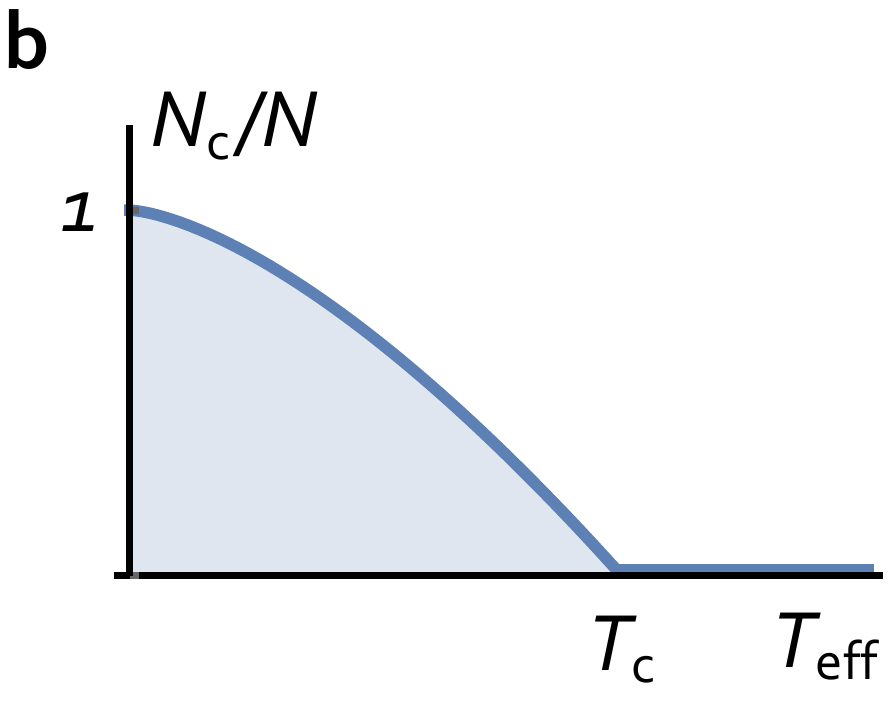} \hskip0.06\linewidth
      \includegraphics[width=0.23\linewidth]{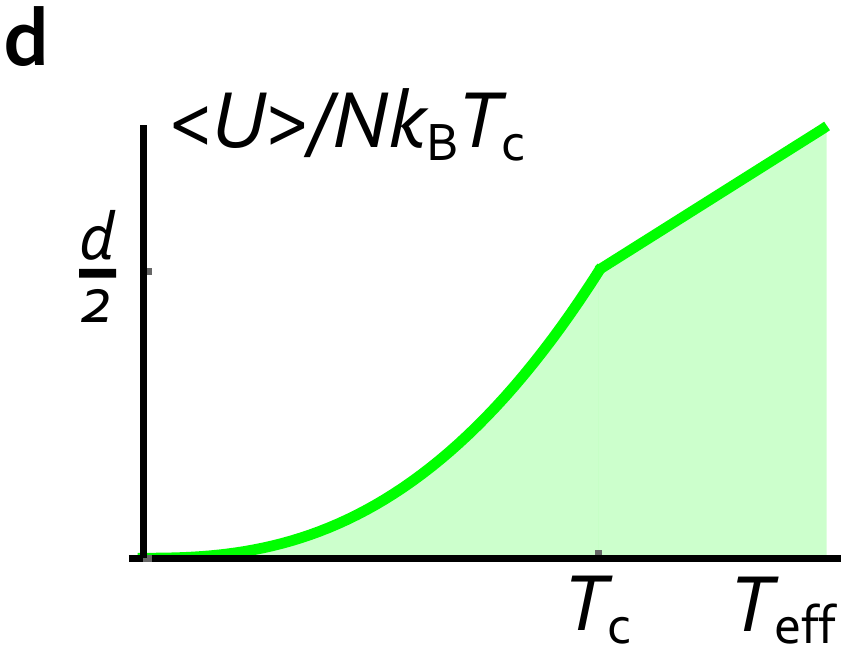} \hskip0.06\linewidth
        \includegraphics[width=0.23\linewidth]{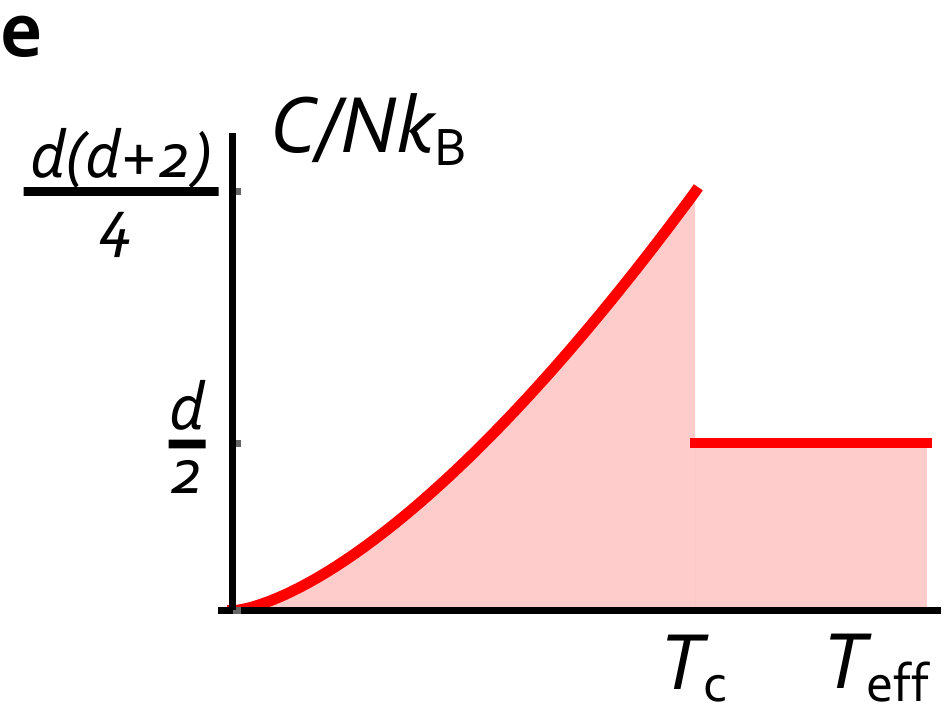} 
 \caption{The effect of inhibition in the presence of a diffusivity edge. (a) The density-dependent diffusivity for an inhibited system where the collective inhibition effects cause the diffusivity to decrease until it vanishes. (b) The fraction of particles in the Bose condensate as a function of the effective temperature near the critical transition temperature $\Tc$. (c) The stationary-state distribution for different values of $\Rc$ vs $\rho_0$. (d) The average internal energy for a $d$ dimensional system with a step function diffusivity profile. (e) The heat capacity of the system showing an overshoot and a discontinuity.\label{fig:inhibited}}
 \end{figure*}

The normalization condition for a total of $N$ particles in $d$-dimensions reads $N=\int \dd^d \rr \rho(U(\rr))=\int S_d r^{d-1} \dd r \rho(U(r))$ assuming that the potential is spherically symmetric, where $S_d=2 \pi^{d/2}/\Gamma(d/2)$ is the surface area of the unit sphere embedded in $d$ dimensions. If $U(r)$ can be inverted (to a single-valued function), the normalization condition can be written as $N= \int_0^{\infty} \dd U g(U) \rho(U)$ where $g(U) \equiv S_d \left(r(U)\right)^{d-1} \frac{\dd r}{\dd U}$ is the density of states. For example, a harmonic trap in the form of $U=\frac{1}{2} k r^2$ yields $g(U)=\frac{S_d}{2} \left(\frac{2}{k}\right)^{d/2} U^{\frac{d}{2}-1}$. Since $U$ is a monotonically decaying function of $\rho$, we can rewrite the normalization condition as follows
\begin{equation}
N=\frac{k_{\rm B}\Te}{\Ds}  \int_{0}^{\rho_0} \dd \rho \,D(\rho) \, g(U(\rho)). \;\;\; (\rho_0 < \Rc) \label{eq:N-1}
\end{equation}
This condition gives $\rho_0(\Te)$, which we expect to be a decreasing function of temperature. However, it only applies when $\rho_0 < \Rc$ as noted. This is because when $\rho_0 \geq \Rc$, the integrand in equation (\ref{eq:N-1}) vanishes identically and no longer contributes, which makes it impossible for the normalization to be satisfied. Therefore, the system develops a condensate with $N_{\rm c}$ particles at the ground state, and the normalization reads
\begin{equation}
N=N_{\rm c}+\frac{k_{\rm B}\Te}{\Ds}  \int_{0}^{\Rc} \dd \rho \,D(\rho) \, g(U(\rho)). \;\;\; (\rho_0 \geq \Rc) \label{eq:N-2}
\end{equation}
We can calculate the size of the condensate as a function of temperature below the transition temperature $\Tc$, which can be obtained by setting $\rho_0(\Tc)=\Rc$. When $\Te \leq \Tc$ (corresponding to $\rho_0 \geq \Rc$), equation (\ref{eq:U-1}) tells us that $\beta U(\rho)=u(\rho/\Rc)$. Noting that $D(\rho)=\Ds \gamma(\rho/\Rc)$ by definition, we can recast equation (\ref{eq:N-2}) into the form $N=N_{\rm c}+(k_{\rm B} \Te)^{d/2}\cdot \frac{S_d}{2} \left(\frac{2}{k}\right)^{d/2} \Rc \int_{0}^{1} \dd s \,\gamma(s) \left[u(s)\right]^{\frac{d}{2}-1}$. Noting that $\Tc$ corresponds to $N_{\rm c}=0$, this calculations yields the fraction of particles in the condensate as
\begin{equation}
\frac{\Nc}{N}=1-\left(\frac{\Te}{\Tc}\right)^{d/2}. \label{eq:NcN}
\end{equation}
This result is plotted in Fig. \ref{fig:inhibited}b. Below, we will examine the behaviour of systems that undergo such a condensation transition more concretely, using a number of specific cases.

 \begin{figure*}[t]
 \includegraphics[width=0.31\linewidth]{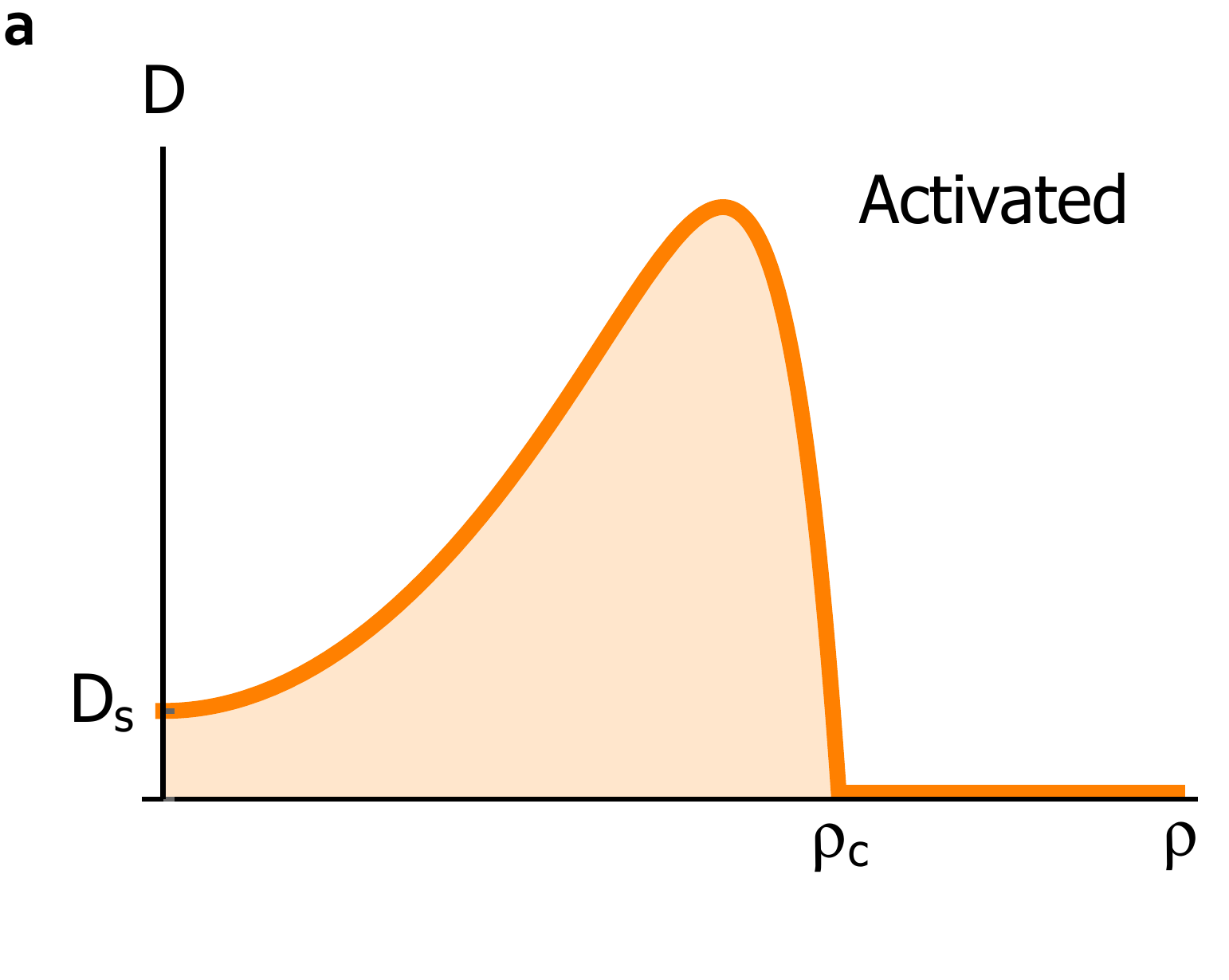} \hskip0.06\linewidth
  \includegraphics[width=0.18\linewidth]{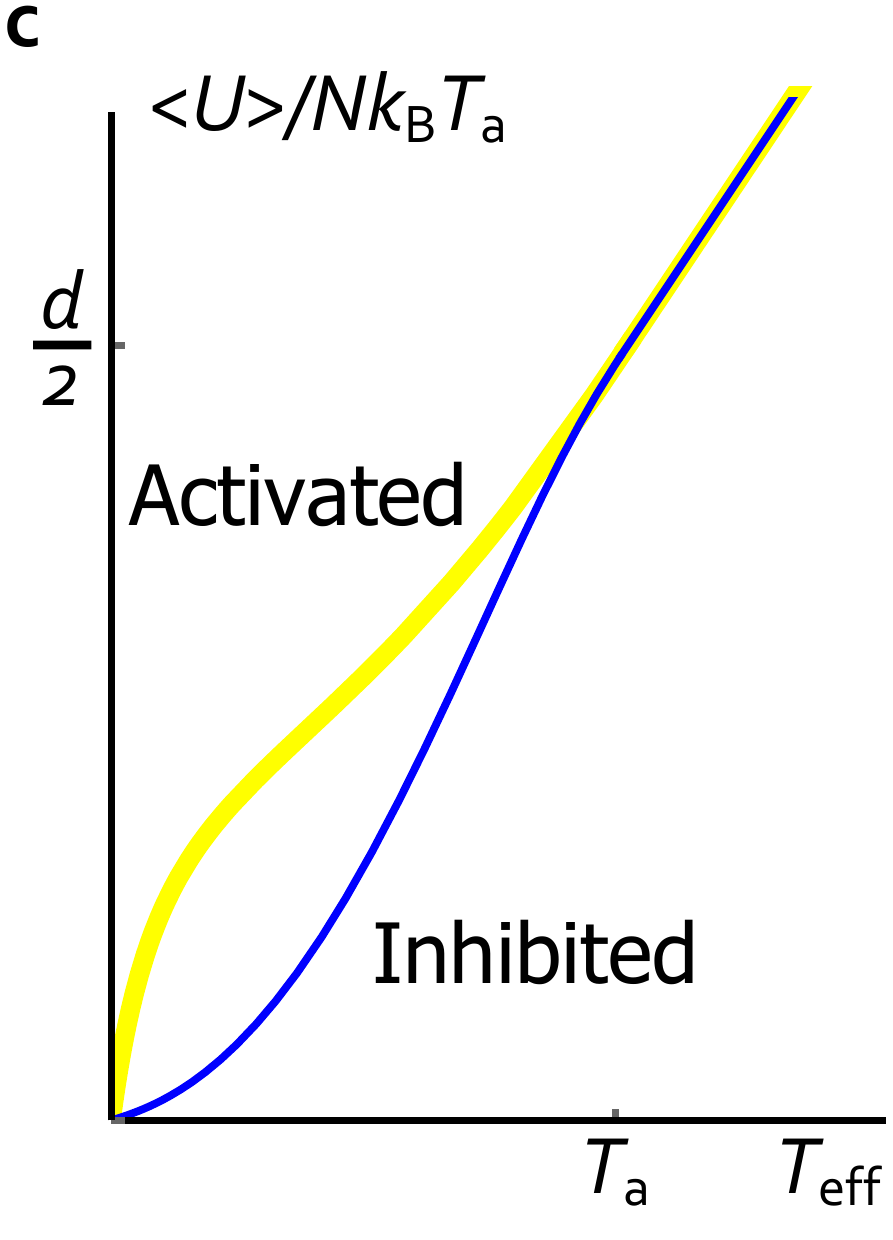} \hskip0.06\linewidth
    \includegraphics[width=0.18\linewidth]{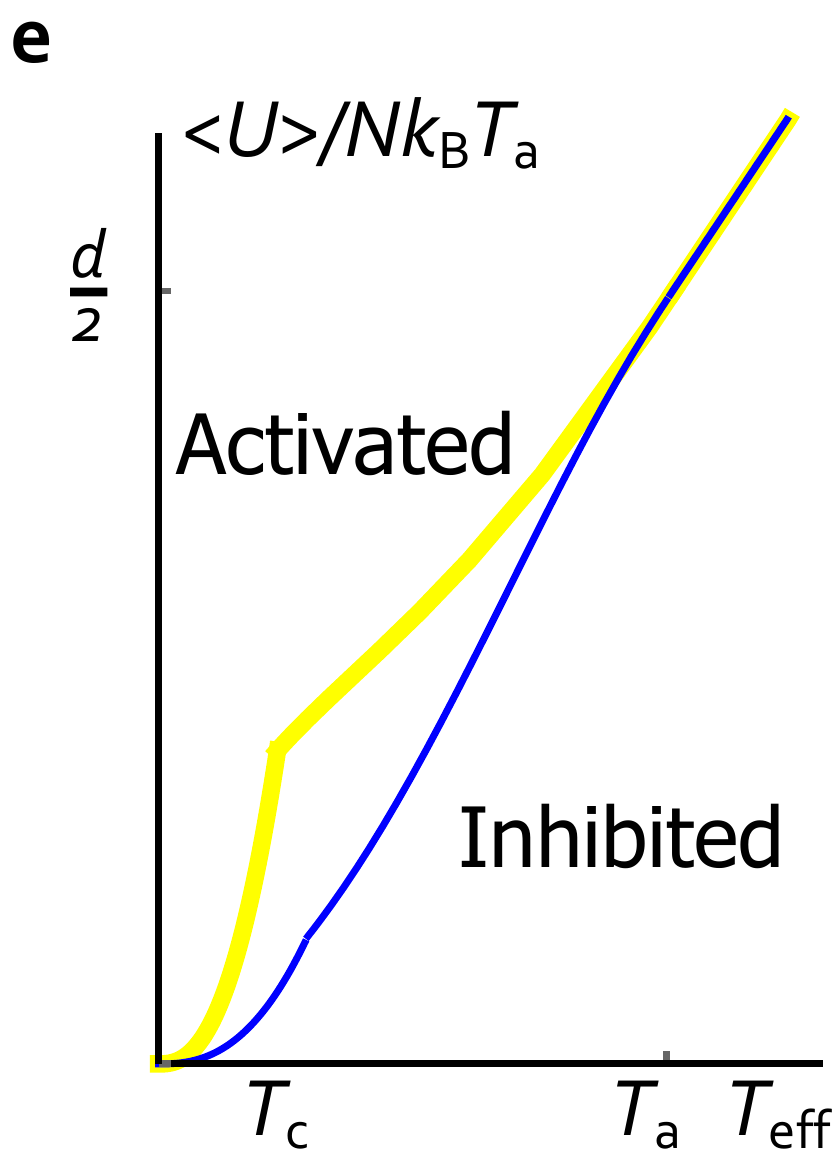} 
  \vskip5mm
      \includegraphics[width=0.23\linewidth]{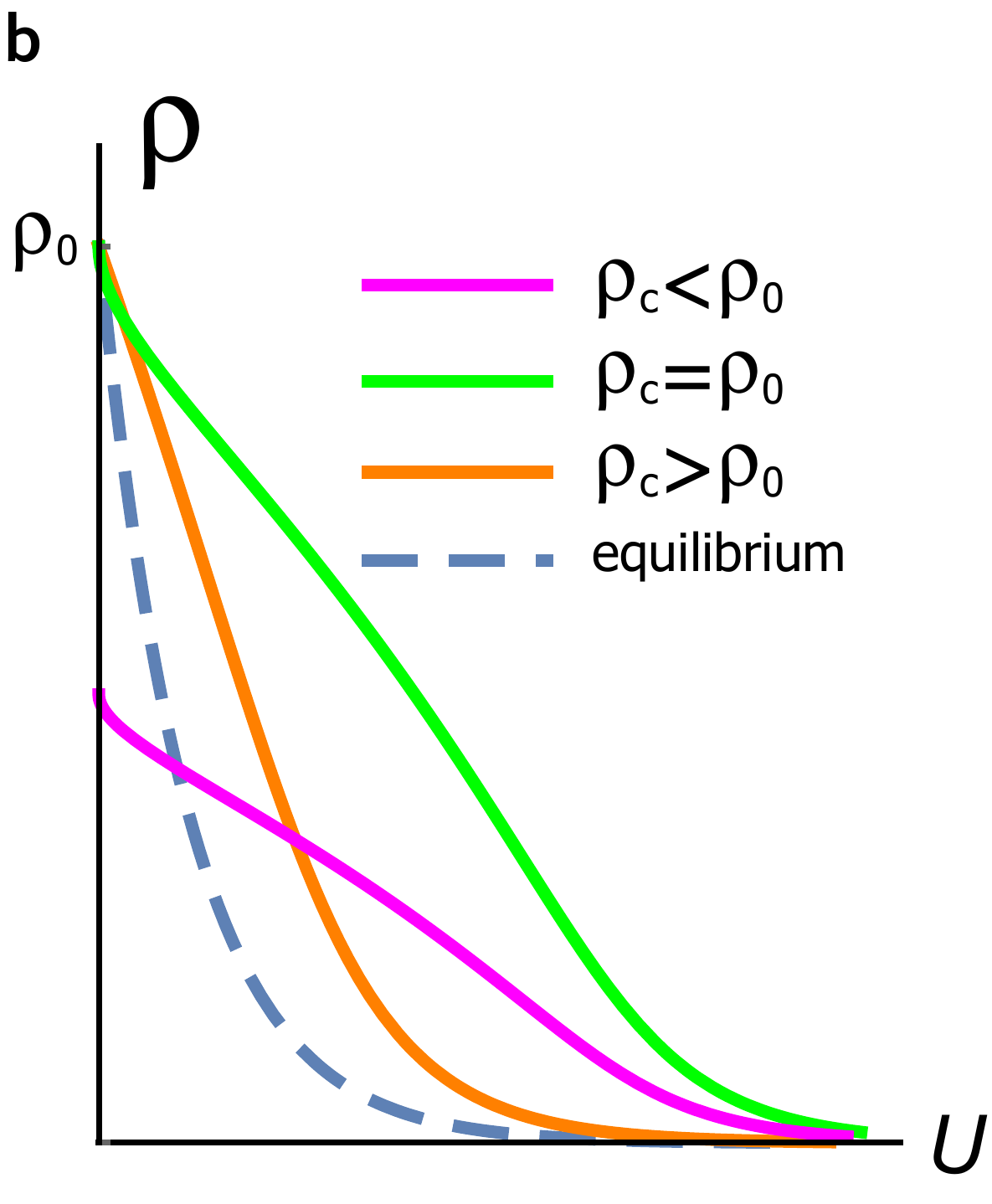} \hskip0.06\linewidth
        \includegraphics[width=0.22\linewidth]{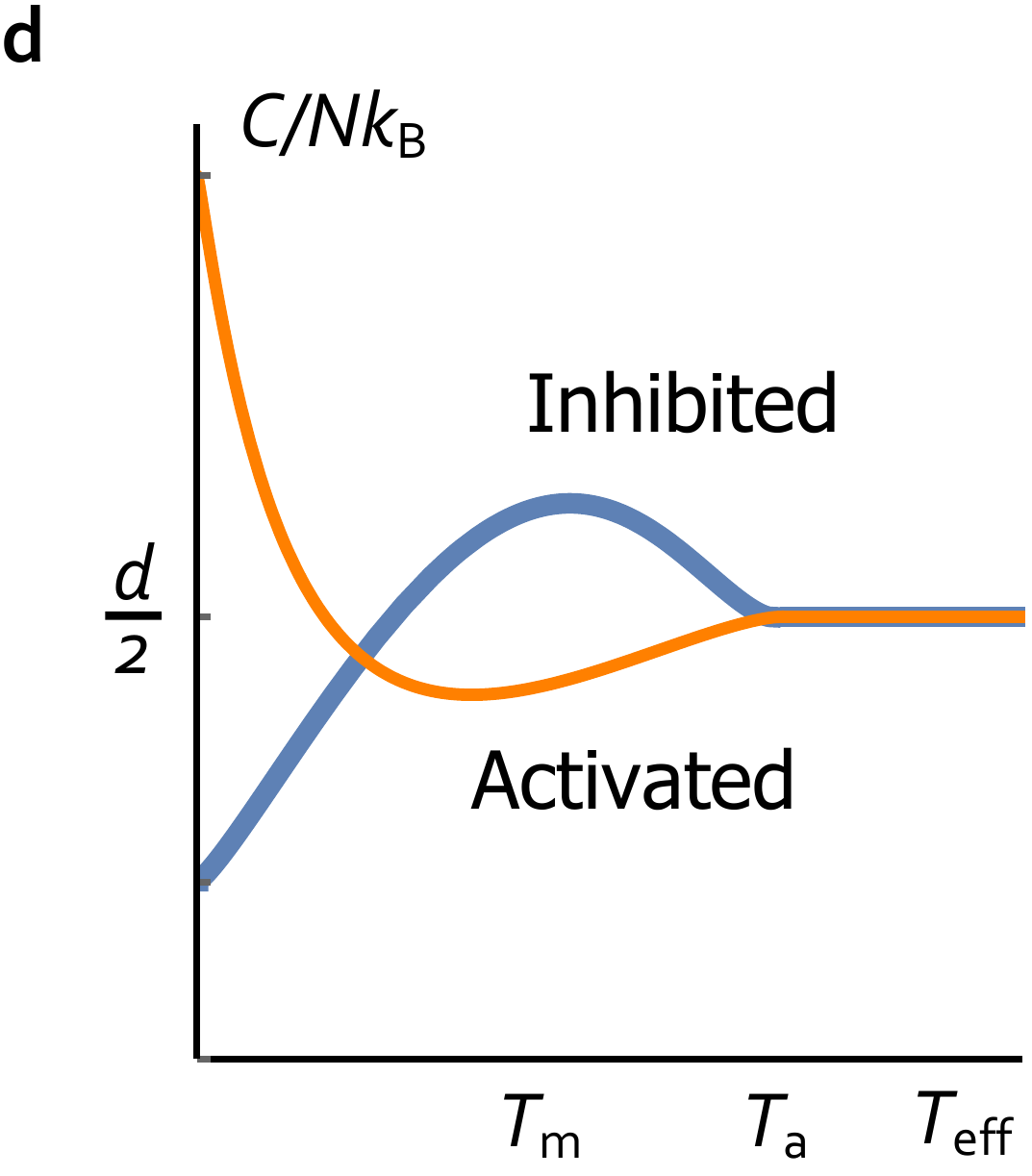} \hskip0.06\linewidth
         \includegraphics[width=0.22\linewidth]{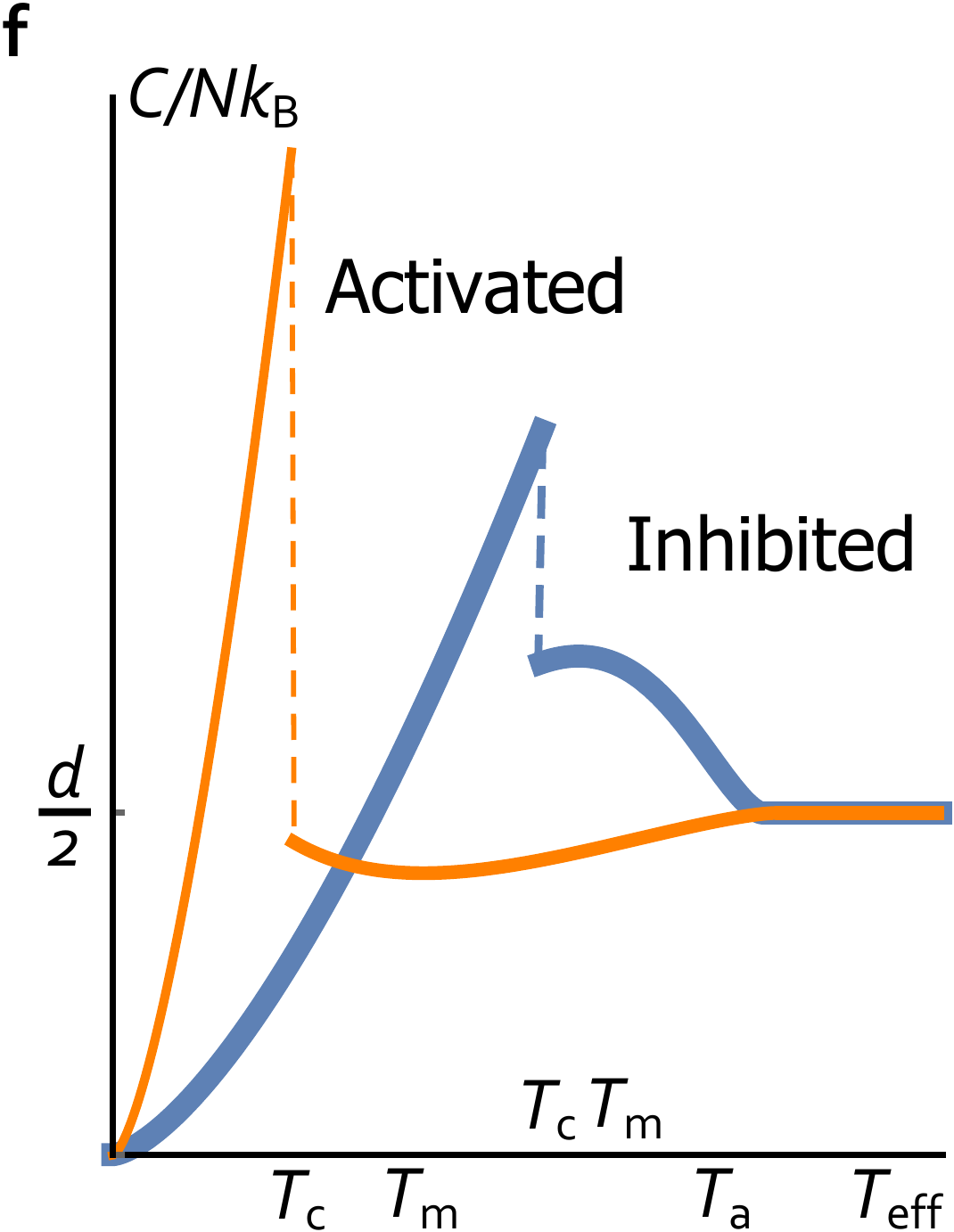} 
 \caption{The effect of activation in the presence a diffusivity edge. (a) The density-dependent diffusivity for an activated system where motility causes the diffusivity to increase initially until it starts to rapidly decrease due to inhibition and finally vanishes. (b) The resulting stationary-state distribution for different values of $\Rc$ vs $\rho_0$. (c) The average internal energy for a $d$ dimensional system as a function of temperature near the activity temperature $\Ta$, for $\Rc \to \infty$ (corresponding to $\Tc \to 0$). The activated case corresponds to $\alpha >1$ and the inhibited case corresponds to $\alpha <1$. (d) The heat capacity of the system showing a dip for the activated case and an overshoot for the inhibited case, at an intermediate temperature $T_{\rm m} < \Ta$, for $\Rc \to \infty$ (corresponding to $\Tc \to 0$). The intercept at $\Te=0$ is equal to $\alpha d/2$. (e) The average internal energy for finite $\Rc$, where a BEC appears at $\Te < \Tc$. (f) The heat capacity for the case with finite $\Rc$. The plot shows an example where $\Tc < T_{\rm m}$. It is also possible to have the condensation before the temperature reaches the maximum or minimum. \label{fig:active}}
 \end{figure*}

Consider a situation where inhibition causes the effective diffusivity in this mean-field description to monotonically decrease as a function of density, until it vanishes at $\rho=\Rc$. Figure \ref{fig:inhibited}a shows the density-dependent diffusivity, and the resulting stationary distribution $\rho(U)$ is presented in Fig. \ref{fig:inhibited}c. For $\rho_0 \ll \Rc $ the distribution approaches a Boltzmann weight $\rho_0 e^{-\beta U}$ while increasing $\rho_0$ with respect to $\Rc$ leads to progressively faster-than-exponential decay, until the slope of the distribution diverges at $U=0$ for $\rho_0=\Rc$, which is the onset of condensation. For $\rho_0 > \Rc$, the slope at $U=0$ continues to diverge, while the asymptotic value $\rho(U \to 0^+)=\Rc$ will be systematically smaller than $\rho_0$. This signals the presence of a condensate at $U=0$, which populates $N_{\rm c}$ particles as determined by equations (\ref{eq:N-2}) and (\ref{eq:NcN}).

To better illustrate the properties of the condensation transition let us consider a step function profile for diffusivity, namely, $D(\rho < \Rc)=\Ds$ and $D(\rho > \Rc)=0$. Then, for the stationary distribution $\rho$ we find $\rho_0 \, e^{-\beta U}$ for $\rho_0 < \Rc$ and $ \Rc \, e^{-\beta U}$ for $\rho_0 \geq \Rc$. Using this explicit form, we can use the normalization condition equation (\ref{eq:N-1}) to obtain the value of $\rho_0$, which for the harmonic potential yields $\rho_0={N}{\left(2 \pi k_{\rm B} \Te/k\right)^{-d/2}}$. Using this, we can find the transition temperature as $\Tc=\frac{k}{2 \pi k_{\rm B}} \left({N}/{\Rc}\right)^{2/d}$ by setting $\rho_0(\Tc)=\Rc$. We can also calculate the average energy of the system as $\left\langle U \right\rangle= \int_0^{\infty} \dd U g(U) U \rho(U)$. This yields
\begin{equation}\label{eq:U-ave-1}
\left\langle U \right\rangle=\left.
  \begin{cases}
   \frac{d}{2}  \cdot N \cdot  k_{\rm B} \Te , & \text{for } \rho_0 < \Rc, \\
   \frac{d}{2} \left(\frac{2 \pi}{k}\right)^{d/2} \Rc \left( k_{\rm B} \Te\right)^{\frac{d}{2}+1}, & \text{for } \rho_0 \geq \Rc, 
  \end{cases}\right.
\end{equation}
which can be rewritten as
\begin{equation}\label{eq:U-ave-2}
\left\langle U \right\rangle=\frac{d}{2}  \cdot N  \cdot k_{\rm B} \Te  \left.
  \begin{cases}
   \left(\frac{\Te}{\Tc}\right)^{d/2} , & \text{for } \Te \leq \Tc, \\
  1, & \text{for } \Te > \Tc.
  \end{cases}\right.
\end{equation}
This result is plotted in Fig. \ref{fig:inhibited}d. Consequently, we obtain the following expression for the heat capacity of the system
\begin{equation}\label{eq:C-1}
C=\frac{\dd \left\langle U \right\rangle}{\dd \Te}=  \frac{d}{2}  \cdot N  k_{\rm B} \left.
  \begin{cases}
   \left(\frac{d}{2}+1\right)  \left(\frac{\Te}{\Tc}\right)^{d/2} , & \text{for } \Te \leq \Tc, \\
  1, & \text{for } \Te > \Tc,
  \end{cases}\right.
\end{equation}
which is plotted in Fig. \ref{fig:inhibited}e. Equations (\ref{eq:NcN}) and (\ref{eq:C-1}) highlight a strong analogy to Bose--Einstein condensation\cite{London:1938,kardar2007statistical}.

We next consider a $D(\rho)$ profile that contains collective activation at intermediate densities before the inhibition at higher level of crowding gives rise to a diffusivity edge, as shown in Fig. \ref{fig:active}a. In this case, the stationary distribution that is shown in Fig.  \ref{fig:active}b shows a tendency for the particles to occupy higher energy states more than the equilibrium case, while the formation of the Bose condensate happens by following the same stages as in the inhibited case, namely, divergence of the slope of $\rho(U)$ at $U=0$ when $\rho_0=\Rc$ and the subsequent depletion that is accompanied by the formation of the condensate at the ground state. 

To further examine the effect of activity, we consider a piecewise diffusivity profile
 \begin{equation}\label{eq:Y-active-1}
D(\rho)=\left.
  \begin{cases}
   \Ds , & \text{for } 0 \leq\rho < \Ra, \\
   \Da , & \text{for } \Ra \leq \rho < \Rc, \\
  0, & \text{for }  \Rc \leq \rho, 
  \end{cases}\right.
\end{equation}
where $\Da$ is the diffusivity in the activated region. Let us define $\alpha\equiv \Da/\Ds$ as the measure of activation. The behaviour of the system will depend on how the ground-state density $\rho_0$ compares with the density scales $\Ra$ and $\Rc$. The different categories are discussed below.

\paragraph{Low density regime $\rho_0 < \Ra$.}The stationary solution in this regime is found as the Boltzmann weight $\rho_0 e^{-\beta U}$ with  $\rho_0={N}{\left(2 \pi k_{\rm B} \Te/k\right)^{-d/2}}$ and the average energy given as $\left\langle U \right\rangle=\frac{d}{2}  \cdot N  \cdot k_{\rm B} \Te$, which yields $C= \frac{d}{2} \cdot N  k_{\rm B}$. Setting $\rho_0=\Ra$, we find the transition temperature $\Ta=\frac{k}{2 \pi k_{\rm B}} \left({N}/{\Ra}\right)^{2/d}$ below which the system will start to be influenced by the activation. The dilute regime corresponds to $\Te > \Ta$.

\paragraph{Intermediate density regime $\Ra \leq \rho_0 < \Rc$.}In this range of densities, which corresponds to $\Te \leq \Ta$, the stationary solution is found as follows
\begin{equation}\label{eq:rho-active-1}
\rho(U)=\left.
  \begin{cases}
  \Ra \left({\rho_0}/{\Ra}\right)^\alpha e^{-\beta U} , & \text{for } \rho < \Ra, \\
  \rho_0 \, e^{-\beta U/\alpha}, & \text{for } \Ra \leq \rho < \rho_0, \\
  \end{cases}\right.
\end{equation}
which resembles a distribution with two different temperatures for different ranges of energy. In this regime, we obtain the particle number normalization condition as
\begin{equation}
N= \frac{\Ra \left({2 \pi}/{k}\right)^{d/2}}{\Gamma(d/2)} \left( k_{\rm B} \Te\right)^{\frac{d}{2}} H_{\frac{d}{2}}\left(\alpha,\ln \frac{\rho_0}{\Ra}\right),\label{eq:N-active-1}
\end{equation}
and the average internal energy content as
\begin{equation}
 \left\langle U \right\rangle= \frac{\Ra \left({2 \pi}/{k}\right)^{d/2}}{\Gamma(d/2)} \left( k_{\rm B} \Te\right)^{\frac{d}{2}+1} H_{\frac{d}{2}+1}\left(\alpha,\ln \frac{\rho_0}{\Ra}\right),\label{eq:U-active-1}
\end{equation}
in terms of the following function
\begin{equation}
H_z(\alpha,p) \equiv e^p \alpha^z \left(\Gamma(z)-\Gamma(z,p)\right)+e^{p \alpha} \, \Gamma(z,\alpha p), \label{eq:Hz-def}
\end{equation}
where $\Gamma(z,p)=\int_p^\infty \dd x \, x^{z-1} e^{-x}$ is the upper incomplete gamma function. From the expression for average energy, we can calculate the heat capacity as follows
\begin{eqnarray}
&&\frac{C}{N k_{\rm B}}=\left(\frac{d}{2}+1\right) \frac{H_{\frac{d}{2}+1}\left(\alpha,\ln \frac{\rho_0}{\Ra}\right)}{H_{\frac{d}{2}}\left(\alpha,\ln \frac{\rho_0}{\Ra}\right)} \label{eq:C-active-1} \\
&&-\left(\frac{d}{2}\right) \frac{H_{\frac{d}{2}+1}\left(\alpha,\ln \frac{\rho_0}{\Ra}\right)+(\alpha-1) \left(\frac{\rho_0}{\Ra}\right)^\alpha \Gamma(\frac{d}{2}+1,\alpha \ln \frac{\rho_0}{\Ra})}{H_{\frac{d}{2}}\left(\alpha,\ln \frac{\rho_0}{\Ra}\right)+(\alpha-1) \left(\frac{\rho_0}{\Ra}\right)^\alpha \Gamma(\frac{d}{2},\alpha \ln \frac{\rho_0}{\Ra})}. \nonumber 
\end{eqnarray}
The average internal energy (shown in Fig. \ref{fig:active}c) starts with a negative curvature for activated particles ($\alpha> 1$) and a positive curvature for inhibited particles ($\alpha < 1$). The slope of this curve corresponds to the heat capacity, which is shown in Fig. \ref{fig:active}d. It starts from $\alpha d/2$ (in units of $N k_{\rm B}$) at small effective temperatures and goes through a ripple before asymptotically approaching $d/2$ at sufficiently large effective temperatures. These results are valid at all effective temperatures (or densities) when $\Rc \to \infty$.

\paragraph{High density regime  $\Rc \leq \rho_0$.}For finite $\Rc$ there exists a regime at which the condensation happens when the density surpasses $\Rc$. The stationary distribution in this case, which corresponds to $\Te \leq \Tc < \Ta$, is given as
\begin{equation}\label{eq:rho-active-2}
\rho(U)=\left.
  \begin{cases}
  \Ra \left({\Rc}/{\Ra}\right)^\alpha e^{-\beta U} , & \text{for } \rho < \Ra, \\
  \Rc \, e^{-\beta U/\alpha}, & \text{for } \Ra \leq \rho < \Rc, \\
  \text{condensate},  & \text{for } \Rc \leq \rho < \rho_0, \\
  \end{cases}\right.
\end{equation}
 using which we can calculate the average internal energy as
\begin{equation}
 \left\langle U \right\rangle=N \cdot  \frac{H_{\frac{d}{2}+1}\left(\alpha,\ln \frac{\Rc}{\Ra}\right)}{H_{\frac{d}{2}}\left(\alpha,\ln \frac{\Rc}{\Ra}\right)} \left(\frac{\Te}{\Tc}\right)^{d/2} k_{\rm B} \Te,\label{eq:U-active-2}
\end{equation}
and the heat capacity of the system as
\begin{equation}
C=N k_{\rm B} \left(\frac{d}{2}+1\right)  \frac{H_{\frac{d}{2}+1}\left(\alpha,\ln \frac{\Rc}{\Ra}\right)}{H_{\frac{d}{2}}\left(\alpha,\ln \frac{\Rc}{\Ra}\right)} \left(\frac{\Te}{\Tc}\right)^{d/2}.  \label{eq:C-active-2}
\end{equation}
The corresponding plots of energy and heat capacity in this case are given in Fig. \ref{fig:active}e and Fig. \ref{fig:active}f, respectively, where the system develops the condensate at $\Te \leq \Tc$, with the resulting characteristic discontinuity in the heat capacity.

We have thus shown that the existence of a diffusivity edge leads to the formation of a condensate, through a transition that is formally equivalent to Bose-Einstein condensation. Trapping scalar active matter allows us to manipulate it using the effective temperature that can be extracted from the ratio between the asymptotic values of the diffusivity and the mobility at the dilute limit, to observe the variety of signatures that exist in the generalized thermodynamic quantities as described above, and in particular, the heat capacity.

Our work has similarities with the formulation that is used to describe motility-induced phase separation (MIPS), with the key difference that the instability in MIPS is triggered by the effective diffusivity changing sign thereby promoting the formation of a dense cluster beyond a threshold density\cite{Cates:2015}. Enforcing the diffusivity edge preempts that instability and gives rise to a new universality class that is formally equivalent to a BEC. We also note that in quantum mechanics we are limited to very specific forms of BEC, as the kinetic energy of bosonic particles can adopt only a limited number of forms; e.g. $p^2/2m$ for non-relativistic massive bosons in dilute gases. Our formulation allows us to explore countless new classes of BEC by designing appropriate forms of external trapping potential, say by using holographic optical traps\cite{Bechinger:2016}. Such generalizations will have a remarkable prospect, as they might unravel new non-equilibrium physics -- of the type that generalized quantum correlated systems might some day be able to reproduce -- using synthetic active matter.

\bibliography{master-bib}

\begin{thebibliography}{30}%
\makeatletter
\providecommand \@ifxundefined [1]{%
 \@ifx{#1\undefined}
}%
\providecommand \@ifnum [1]{%
 \ifnum #1\expandafter \@firstoftwo
 \else \expandafter \@secondoftwo
 \fi
}%
\providecommand \@ifx [1]{%
 \ifx #1\expandafter \@firstoftwo
 \else \expandafter \@secondoftwo
 \fi
}%
\providecommand \natexlab [1]{#1}%
\providecommand \enquote  [1]{``#1''}%
\providecommand \bibnamefont  [1]{#1}%
\providecommand \bibfnamefont [1]{#1}%
\providecommand \citenamefont [1]{#1}%
\providecommand \href@noop [0]{\@secondoftwo}%
\providecommand \href [0]{\begingroup \@sanitize@url \@href}%
\providecommand \@href[1]{\@@startlink{#1}\@@href}%
\providecommand \@@href[1]{\endgroup#1\@@endlink}%
\providecommand \@sanitize@url [0]{\catcode `\\12\catcode `\$12\catcode
  `\&12\catcode `\#12\catcode `\^12\catcode `\_12\catcode `\%12\relax}%
\providecommand \@@startlink[1]{}%
\providecommand \@@endlink[0]{}%
\providecommand \url  [0]{\begingroup\@sanitize@url \@url }%
\providecommand \@url [1]{\endgroup\@href {#1}{\urlprefix }}%
\providecommand \urlprefix  [0]{URL }%
\providecommand \Eprint [0]{\href }%
\providecommand \doibase [0]{http://dx.doi.org/}%
\providecommand \selectlanguage [0]{\@gobble}%
\providecommand \bibinfo  [0]{\@secondoftwo}%
\providecommand \bibfield  [0]{\@secondoftwo}%
\providecommand \translation [1]{[#1]}%
\providecommand \BibitemOpen [0]{}%
\providecommand \bibitemStop [0]{}%
\providecommand \bibitemNoStop [0]{.\EOS\space}%
\providecommand \EOS [0]{\spacefactor3000\relax}%
\providecommand \BibitemShut  [1]{\csname bibitem#1\endcsname}%
\let\auto@bib@innerbib\@empty
\bibitem [{\citenamefont {Marchetti}\ \emph {et~al.}(2013)\citenamefont
  {Marchetti}, \citenamefont {Joanny}, \citenamefont {Ramaswamy}, \citenamefont
  {Liverpool}, \citenamefont {Prost}, \citenamefont {Rao},\ and\ \citenamefont
  {Simha}}]{marchetti2013hydrodynamics}%
  \BibitemOpen
  \bibfield  {author} {\bibinfo {author} {\bibfnamefont {M.~C.}\ \bibnamefont
  {Marchetti}}, \bibinfo {author} {\bibfnamefont {J.~F.}\ \bibnamefont
  {Joanny}}, \bibinfo {author} {\bibfnamefont {S.}~\bibnamefont {Ramaswamy}},
  \bibinfo {author} {\bibfnamefont {T.~B.}\ \bibnamefont {Liverpool}}, \bibinfo
  {author} {\bibfnamefont {J.}~\bibnamefont {Prost}}, \bibinfo {author}
  {\bibfnamefont {Madan}\ \bibnamefont {Rao}}, \ and\ \bibinfo {author}
  {\bibfnamefont {R.~Aditi}\ \bibnamefont {Simha}},\ }\bibfield  {title}
  {\enquote {\bibinfo {title} {Hydrodynamics of soft active matter},}\ }\href
  {\doibase 10.1103/RevModPhys.85.1143} {\bibfield  {journal} {\bibinfo
  {journal} {Rev. Mod. Phys.}\ }\textbf {\bibinfo {volume} {85}},\ \bibinfo
  {pages} {1143--1189} (\bibinfo {year} {2013})}\BibitemShut {NoStop}%
\bibitem [{\citenamefont {Bechinger}\ \emph {et~al.}(2016)\citenamefont
  {Bechinger}, \citenamefont {Di~Leonardo}, \citenamefont {L\"owen},
  \citenamefont {Reichhardt}, \citenamefont {Volpe},\ and\ \citenamefont
  {Volpe}}]{Bechinger:2016}%
  \BibitemOpen
  \bibfield  {author} {\bibinfo {author} {\bibfnamefont {Clemens}\ \bibnamefont
  {Bechinger}}, \bibinfo {author} {\bibfnamefont {Roberto}\ \bibnamefont
  {Di~Leonardo}}, \bibinfo {author} {\bibfnamefont {Hartmut}\ \bibnamefont
  {L\"owen}}, \bibinfo {author} {\bibfnamefont {Charles}\ \bibnamefont
  {Reichhardt}}, \bibinfo {author} {\bibfnamefont {Giorgio}\ \bibnamefont
  {Volpe}}, \ and\ \bibinfo {author} {\bibfnamefont {Giovanni}\ \bibnamefont
  {Volpe}},\ }\bibfield  {title} {\enquote {\bibinfo {title} {Active particles
  in complex and crowded environments},}\ }\href {\doibase
  10.1103/RevModPhys.88.045006} {\bibfield  {journal} {\bibinfo  {journal}
  {Rev. Mod. Phys.}\ }\textbf {\bibinfo {volume} {88}},\ \bibinfo {pages}
  {045006} (\bibinfo {year} {2016})}\BibitemShut {NoStop}%
\bibitem [{\citenamefont {Elgeti}\ \emph {et~al.}(2015)\citenamefont {Elgeti},
  \citenamefont {Winkler},\ and\ \citenamefont {Gompper}}]{Elgeti:2015}%
  \BibitemOpen
  \bibfield  {author} {\bibinfo {author} {\bibfnamefont {J}~\bibnamefont
  {Elgeti}}, \bibinfo {author} {\bibfnamefont {R~G}\ \bibnamefont {Winkler}}, \
  and\ \bibinfo {author} {\bibfnamefont {G}~\bibnamefont {Gompper}},\
  }\bibfield  {title} {\enquote {\bibinfo {title} {Physics of
  microswimmers—single particle motion and collective behavior: a review},}\
  }\href {http://stacks.iop.org/0034-4885/78/i=5/a=056601} {\bibfield
  {journal} {\bibinfo  {journal} {Reports on Progress in Physics}\ }\textbf
  {\bibinfo {volume} {78}},\ \bibinfo {pages} {056601} (\bibinfo {year}
  {2015})}\BibitemShut {NoStop}%
\bibitem [{\citenamefont {Keller}\ and\ \citenamefont
  {Segel}(1971)}]{keller1971model}%
  \BibitemOpen
  \bibfield  {author} {\bibinfo {author} {\bibfnamefont {Evelyn~F.}\
  \bibnamefont {Keller}}\ and\ \bibinfo {author} {\bibfnamefont {Lee~A.}\
  \bibnamefont {Segel}},\ }\bibfield  {title} {\enquote {\bibinfo {title}
  {Model for chemotaxis},}\ }\href {\doibase
  https://doi.org/10.1016/0022-5193(71)90050-6} {\bibfield  {journal} {\bibinfo
   {journal} {Journal of Theoretical Biology}\ }\textbf {\bibinfo {volume}
  {30}},\ \bibinfo {pages} {225 -- 234} (\bibinfo {year} {1971})}\BibitemShut
  {NoStop}%
\bibitem [{\citenamefont {Wadhams}\ and\ \citenamefont
  {Armitage}(2004)}]{wadh04}%
  \BibitemOpen
  \bibfield  {author} {\bibinfo {author} {\bibfnamefont {George~H.}\
  \bibnamefont {Wadhams}}\ and\ \bibinfo {author} {\bibfnamefont {Judith~P.}\
  \bibnamefont {Armitage}},\ }\bibfield  {title} {\enquote {\bibinfo {title}
  {{Making sense of it all: bacterial chemotaxis}},}\ }\href {\doibase
  10.1038/nrm1524} {\bibfield  {journal} {\bibinfo  {journal} {Nat. Rev. Mol.
  Cell Biol.}\ }\textbf {\bibinfo {volume} {5}},\ \bibinfo {pages} {1024--1037}
  (\bibinfo {year} {2004})}\BibitemShut {NoStop}%
\bibitem [{\citenamefont {Taktikos}\ \emph {et~al.}(2012)\citenamefont
  {Taktikos}, \citenamefont {Zaburdaev},\ and\ \citenamefont
  {Stark}}]{taktikos+zaburdaev12}%
  \BibitemOpen
  \bibfield  {author} {\bibinfo {author} {\bibfnamefont {Johannes}\
  \bibnamefont {Taktikos}}, \bibinfo {author} {\bibfnamefont {Vasily}\
  \bibnamefont {Zaburdaev}}, \ and\ \bibinfo {author} {\bibfnamefont {Holger}\
  \bibnamefont {Stark}},\ }\bibfield  {title} {\enquote {\bibinfo {title}
  {Collective dynamics of model microorganisms with chemotactic signaling},}\
  }\href {\doibase 10.1103/PhysRevE.85.051901} {\bibfield  {journal} {\bibinfo
  {journal} {Phys. Rev. E}\ }\textbf {\bibinfo {volume} {85}},\ \bibinfo
  {pages} {051901} (\bibinfo {year} {2012})}\BibitemShut {NoStop}%
\bibitem [{\citenamefont {Saha}\ \emph {et~al.}(2014)\citenamefont {Saha},
  \citenamefont {Golestanian},\ and\ \citenamefont
  {Ramaswamy}}]{saha+golestanian14}%
  \BibitemOpen
  \bibfield  {author} {\bibinfo {author} {\bibfnamefont {Suropriya}\
  \bibnamefont {Saha}}, \bibinfo {author} {\bibfnamefont {Ramin}\ \bibnamefont
  {Golestanian}}, \ and\ \bibinfo {author} {\bibfnamefont {Sriram}\
  \bibnamefont {Ramaswamy}},\ }\bibfield  {title} {\enquote {\bibinfo {title}
  {Clusters, asters, and collective oscillations in chemotactic colloids},}\
  }\href {\doibase 10.1103/PhysRevE.89.062316} {\bibfield  {journal} {\bibinfo
  {journal} {Phys. Rev. E}\ }\textbf {\bibinfo {volume} {89}},\ \bibinfo
  {pages} {062316} (\bibinfo {year} {2014})}\BibitemShut {NoStop}%
\bibitem [{\citenamefont {Sokolov}\ \emph {et~al.}(2007)\citenamefont
  {Sokolov}, \citenamefont {Aranson}, \citenamefont {Kessler},\ and\
  \citenamefont {Goldstein}}]{Sokolov:2007}%
  \BibitemOpen
  \bibfield  {author} {\bibinfo {author} {\bibfnamefont {Andrey}\ \bibnamefont
  {Sokolov}}, \bibinfo {author} {\bibfnamefont {Igor~S.}\ \bibnamefont
  {Aranson}}, \bibinfo {author} {\bibfnamefont {John~O.}\ \bibnamefont
  {Kessler}}, \ and\ \bibinfo {author} {\bibfnamefont {Raymond~E.}\
  \bibnamefont {Goldstein}},\ }\bibfield  {title} {\enquote {\bibinfo {title}
  {Concentration dependence of the collective dynamics of swimming bacteria},}\
  }\href {\doibase 10.1103/PhysRevLett.98.158102} {\bibfield  {journal}
  {\bibinfo  {journal} {Phys. Rev. Lett.}\ }\textbf {\bibinfo {volume} {98}},\
  \bibinfo {pages} {158102} (\bibinfo {year} {2007})}\BibitemShut {NoStop}%
\bibitem [{\citenamefont {Ishikawa}\ \emph {et~al.}(2011)\citenamefont
  {Ishikawa}, \citenamefont {Yoshida}, \citenamefont {Ueno}, \citenamefont
  {Wiedeman}, \citenamefont {Imai},\ and\ \citenamefont
  {Yamaguchi}}]{Ishikawa:2011}%
  \BibitemOpen
  \bibfield  {author} {\bibinfo {author} {\bibfnamefont {T.}~\bibnamefont
  {Ishikawa}}, \bibinfo {author} {\bibfnamefont {N.}~\bibnamefont {Yoshida}},
  \bibinfo {author} {\bibfnamefont {H.}~\bibnamefont {Ueno}}, \bibinfo {author}
  {\bibfnamefont {M.}~\bibnamefont {Wiedeman}}, \bibinfo {author}
  {\bibfnamefont {Y.}~\bibnamefont {Imai}}, \ and\ \bibinfo {author}
  {\bibfnamefont {T.}~\bibnamefont {Yamaguchi}},\ }\bibfield  {title} {\enquote
  {\bibinfo {title} {Energy transport in a concentrated suspension of
  bacteria},}\ }\href {\doibase 10.1103/PhysRevLett.107.028102} {\bibfield
  {journal} {\bibinfo  {journal} {Phys. Rev. Lett.}\ }\textbf {\bibinfo
  {volume} {107}},\ \bibinfo {pages} {028102} (\bibinfo {year}
  {2011})}\BibitemShut {NoStop}%
\bibitem [{\citenamefont {Aditi~Simha}\ and\ \citenamefont
  {Ramaswamy}(2002)}]{Simha:2002}%
  \BibitemOpen
  \bibfield  {author} {\bibinfo {author} {\bibfnamefont {R.}~\bibnamefont
  {Aditi~Simha}}\ and\ \bibinfo {author} {\bibfnamefont {Sriram}\ \bibnamefont
  {Ramaswamy}},\ }\bibfield  {title} {\enquote {\bibinfo {title} {Hydrodynamic
  fluctuations and instabilities in ordered suspensions of self-propelled
  particles},}\ }\href {\doibase 10.1103/PhysRevLett.89.058101} {\bibfield
  {journal} {\bibinfo  {journal} {Phys. Rev. Lett.}\ }\textbf {\bibinfo
  {volume} {89}},\ \bibinfo {pages} {058101} (\bibinfo {year}
  {2002})}\BibitemShut {NoStop}%
\bibitem [{\citenamefont {Saintillan}\ and\ \citenamefont
  {Shelley}(2008)}]{Saintillan:2008}%
  \BibitemOpen
  \bibfield  {author} {\bibinfo {author} {\bibfnamefont {David}\ \bibnamefont
  {Saintillan}}\ and\ \bibinfo {author} {\bibfnamefont {Michael~J.}\
  \bibnamefont {Shelley}},\ }\bibfield  {title} {\enquote {\bibinfo {title}
  {Instabilities and pattern formation in active particle suspensions: Kinetic
  theory and continuum simulations},}\ }\href {\doibase
  10.1103/PhysRevLett.100.178103} {\bibfield  {journal} {\bibinfo  {journal}
  {Phys. Rev. Lett.}\ }\textbf {\bibinfo {volume} {100}},\ \bibinfo {pages}
  {178103} (\bibinfo {year} {2008})}\BibitemShut {NoStop}%
\bibitem [{\citenamefont {Golestanian}(2012)}]{Golestanian:2012}%
  \BibitemOpen
  \bibfield  {author} {\bibinfo {author} {\bibfnamefont {Ramin}\ \bibnamefont
  {Golestanian}},\ }\bibfield  {title} {\enquote {\bibinfo {title} {Collective
  behavior of thermally active colloids},}\ }\href {\doibase
  10.1103/PhysRevLett.108.038303} {\bibfield  {journal} {\bibinfo  {journal}
  {Phys. Rev. Lett.}\ }\textbf {\bibinfo {volume} {108}},\ \bibinfo {pages}
  {038303} (\bibinfo {year} {2012})}\BibitemShut {NoStop}%
\bibitem [{\citenamefont {Cates}\ and\ \citenamefont
  {Tailleur}(2015)}]{Cates:2015}%
  \BibitemOpen
  \bibfield  {author} {\bibinfo {author} {\bibfnamefont {Michael~E.}\
  \bibnamefont {Cates}}\ and\ \bibinfo {author} {\bibfnamefont {Julien}\
  \bibnamefont {Tailleur}},\ }\bibfield  {title} {\enquote {\bibinfo {title}
  {Motility-induced phase separation},}\ }\href {\doibase
  10.1146/annurev-conmatphys-031214-014710} {\bibfield  {journal} {\bibinfo
  {journal} {Annual Review of Condensed Matter Physics}\ }\textbf {\bibinfo
  {volume} {6}},\ \bibinfo {pages} {219--244} (\bibinfo {year} {2015})},\
  \Eprint
  {http://arxiv.org/abs/https://doi.org/10.1146/annurev-conmatphys-031214-014710}
  {https://doi.org/10.1146/annurev-conmatphys-031214-014710} \BibitemShut
  {NoStop}%
\bibitem [{\citenamefont {Henkes}\ \emph {et~al.}(2011)\citenamefont {Henkes},
  \citenamefont {Fily},\ and\ \citenamefont {Marchetti}}]{Henkes:2011}%
  \BibitemOpen
  \bibfield  {author} {\bibinfo {author} {\bibfnamefont {Silke}\ \bibnamefont
  {Henkes}}, \bibinfo {author} {\bibfnamefont {Yaouen}\ \bibnamefont {Fily}}, \
  and\ \bibinfo {author} {\bibfnamefont {M.~Cristina}\ \bibnamefont
  {Marchetti}},\ }\bibfield  {title} {\enquote {\bibinfo {title} {Active
  jamming: Self-propelled soft particles at high density},}\ }\href {\doibase
  10.1103/PhysRevE.84.040301} {\bibfield  {journal} {\bibinfo  {journal} {Phys.
  Rev. E}\ }\textbf {\bibinfo {volume} {84}},\ \bibinfo {pages} {040301}
  (\bibinfo {year} {2011})}\BibitemShut {NoStop}%
\bibitem [{\citenamefont {Redner}\ \emph {et~al.}(2013)\citenamefont {Redner},
  \citenamefont {Hagan},\ and\ \citenamefont {Baskaran}}]{Hagan:2013}%
  \BibitemOpen
  \bibfield  {author} {\bibinfo {author} {\bibfnamefont {Gabriel~S.}\
  \bibnamefont {Redner}}, \bibinfo {author} {\bibfnamefont {Michael~F.}\
  \bibnamefont {Hagan}}, \ and\ \bibinfo {author} {\bibfnamefont {Aparna}\
  \bibnamefont {Baskaran}},\ }\bibfield  {title} {\enquote {\bibinfo {title}
  {Structure and dynamics of a phase-separating active colloidal fluid},}\
  }\href {\doibase 10.1103/PhysRevLett.110.055701} {\bibfield  {journal}
  {\bibinfo  {journal} {Phys. Rev. Lett.}\ }\textbf {\bibinfo {volume} {110}},\
  \bibinfo {pages} {055701} (\bibinfo {year} {2013})}\BibitemShut {NoStop}%
\bibitem [{\citenamefont {Buttinoni}\ \emph {et~al.}(2013)\citenamefont
  {Buttinoni}, \citenamefont {Bialk\'e}, \citenamefont {K\"ummel},
  \citenamefont {L\"owen}, \citenamefont {Bechinger},\ and\ \citenamefont
  {Speck}}]{Buttinoni:2013}%
  \BibitemOpen
  \bibfield  {author} {\bibinfo {author} {\bibfnamefont {Ivo}\ \bibnamefont
  {Buttinoni}}, \bibinfo {author} {\bibfnamefont {Julian}\ \bibnamefont
  {Bialk\'e}}, \bibinfo {author} {\bibfnamefont {Felix}\ \bibnamefont
  {K\"ummel}}, \bibinfo {author} {\bibfnamefont {Hartmut}\ \bibnamefont
  {L\"owen}}, \bibinfo {author} {\bibfnamefont {Clemens}\ \bibnamefont
  {Bechinger}}, \ and\ \bibinfo {author} {\bibfnamefont {Thomas}\ \bibnamefont
  {Speck}},\ }\bibfield  {title} {\enquote {\bibinfo {title} {Dynamical
  clustering and phase separation in suspensions of self-propelled colloidal
  particles},}\ }\href {\doibase 10.1103/PhysRevLett.110.238301} {\bibfield
  {journal} {\bibinfo  {journal} {Phys. Rev. Lett.}\ }\textbf {\bibinfo
  {volume} {110}},\ \bibinfo {pages} {238301} (\bibinfo {year}
  {2013})}\BibitemShut {NoStop}%
\bibitem [{\citenamefont {Soto}\ and\ \citenamefont
  {Golestanian}(2014)}]{soto+golestanian14}%
  \BibitemOpen
  \bibfield  {author} {\bibinfo {author} {\bibfnamefont {Rodrigo}\ \bibnamefont
  {Soto}}\ and\ \bibinfo {author} {\bibfnamefont {Ramin}\ \bibnamefont
  {Golestanian}},\ }\bibfield  {title} {\enquote {\bibinfo {title}
  {Run-and-tumble dynamics in a crowded environment: Persistent exclusion
  process for swimmers},}\ }\href {\doibase 10.1103/PhysRevE.89.012706}
  {\bibfield  {journal} {\bibinfo  {journal} {Phys. Rev. E}\ }\textbf {\bibinfo
  {volume} {89}},\ \bibinfo {pages} {012706} (\bibinfo {year}
  {2014})}\BibitemShut {NoStop}%
\bibitem [{\citenamefont {Blaschke}\ \emph {et~al.}(2016)\citenamefont
  {Blaschke}, \citenamefont {Maurer}, \citenamefont {Menon}, \citenamefont
  {Z\"{o}ttl},\ and\ \citenamefont {Stark}}]{Blaschke:2016}%
  \BibitemOpen
  \bibfield  {author} {\bibinfo {author} {\bibfnamefont {Johannes}\
  \bibnamefont {Blaschke}}, \bibinfo {author} {\bibfnamefont {Maurice}\
  \bibnamefont {Maurer}}, \bibinfo {author} {\bibfnamefont {Karthik}\
  \bibnamefont {Menon}}, \bibinfo {author} {\bibfnamefont {Andreas}\
  \bibnamefont {Z\"{o}ttl}}, \ and\ \bibinfo {author} {\bibfnamefont {Holger}\
  \bibnamefont {Stark}},\ }\bibfield  {title} {\enquote {\bibinfo {title}
  {Phase separation and coexistence of hydrodynamically interacting
  microswimmers},}\ }\href {\doibase 10.1039/c6sm02042a} {\bibfield  {journal}
  {\bibinfo  {journal} {Soft Matter}\ }\textbf {\bibinfo {volume} {12}},\
  \bibinfo {pages} {9821--9831} (\bibinfo {year} {2016})}\BibitemShut {NoStop}%
\bibitem [{\citenamefont {Digregorio}\ \emph {et~al.}(2018)\citenamefont
  {Digregorio}, \citenamefont {Levis}, \citenamefont {Suma}, \citenamefont
  {Cugliandolo}, \citenamefont {Gonnella},\ and\ \citenamefont
  {Pagonabarraga}}]{Letitia:2018}%
  \BibitemOpen
  \bibfield  {author} {\bibinfo {author} {\bibfnamefont {Pasquale}\
  \bibnamefont {Digregorio}}, \bibinfo {author} {\bibfnamefont {Demian}\
  \bibnamefont {Levis}}, \bibinfo {author} {\bibfnamefont {Antonio}\
  \bibnamefont {Suma}}, \bibinfo {author} {\bibfnamefont {Leticia~F.}\
  \bibnamefont {Cugliandolo}}, \bibinfo {author} {\bibfnamefont {Giuseppe}\
  \bibnamefont {Gonnella}}, \ and\ \bibinfo {author} {\bibfnamefont {Ignacio}\
  \bibnamefont {Pagonabarraga}},\ }\bibfield  {title} {\enquote {\bibinfo
  {title} {Full phase diagram of active brownian disks: From melting to
  motility-induced phase separation},}\ }\href {\doibase
  10.1103/PhysRevLett.121.098003} {\bibfield  {journal} {\bibinfo  {journal}
  {Phys. Rev. Lett.}\ }\textbf {\bibinfo {volume} {121}},\ \bibinfo {pages}
  {098003} (\bibinfo {year} {2018})}\BibitemShut {NoStop}%
\bibitem [{\citenamefont {Abaurrea~Velasco}\ \emph {et~al.}(2018)\citenamefont
  {Abaurrea~Velasco}, \citenamefont {Abkenar}, \citenamefont {Gompper},\ and\
  \citenamefont {Auth}}]{Gompper:2018}%
  \BibitemOpen
  \bibfield  {author} {\bibinfo {author} {\bibfnamefont {Clara}\ \bibnamefont
  {Abaurrea~Velasco}}, \bibinfo {author} {\bibfnamefont {Masoud}\ \bibnamefont
  {Abkenar}}, \bibinfo {author} {\bibfnamefont {Gerhard}\ \bibnamefont
  {Gompper}}, \ and\ \bibinfo {author} {\bibfnamefont {Thorsten}\ \bibnamefont
  {Auth}},\ }\bibfield  {title} {\enquote {\bibinfo {title} {Collective
  behavior of self-propelled rods with quorum sensing},}\ }\href {\doibase
  10.1103/PhysRevE.98.022605} {\bibfield  {journal} {\bibinfo  {journal} {Phys.
  Rev. E}\ }\textbf {\bibinfo {volume} {98}},\ \bibinfo {pages} {022605}
  (\bibinfo {year} {2018})}\BibitemShut {NoStop}%
\bibitem [{\citenamefont {Lu}\ \emph {et~al.}(2008)\citenamefont {Lu},
  \citenamefont {Zaccarelli}, \citenamefont {Ciulla}, \citenamefont
  {Schofield}, \citenamefont {Sciortino},\ and\ \citenamefont
  {Weitz}}]{Lu:2008}%
  \BibitemOpen
  \bibfield  {author} {\bibinfo {author} {\bibfnamefont {Peter~J.}\
  \bibnamefont {Lu}}, \bibinfo {author} {\bibfnamefont {Emanuela}\ \bibnamefont
  {Zaccarelli}}, \bibinfo {author} {\bibfnamefont {Fabio}\ \bibnamefont
  {Ciulla}}, \bibinfo {author} {\bibfnamefont {Andrew~B.}\ \bibnamefont
  {Schofield}}, \bibinfo {author} {\bibfnamefont {Francesco}\ \bibnamefont
  {Sciortino}}, \ and\ \bibinfo {author} {\bibfnamefont {David~A.}\
  \bibnamefont {Weitz}},\ }\bibfield  {title} {\enquote {\bibinfo {title}
  {Gelation of particles with short-range attraction},}\ }\href
  {https://doi.org/10.1038/nature06931} {\bibfield  {journal} {\bibinfo
  {journal} {Nature}\ }\textbf {\bibinfo {volume} {453}},\ \bibinfo {pages}
  {499} (\bibinfo {year} {2008})}\BibitemShut {NoStop}%
\bibitem [{\citenamefont {Cavagna}(2009)}]{Cavagna:2009}%
  \BibitemOpen
  \bibfield  {author} {\bibinfo {author} {\bibfnamefont {Andrea}\ \bibnamefont
  {Cavagna}},\ }\bibfield  {title} {\enquote {\bibinfo {title} {Supercooled
  liquids for pedestrians},}\ }\href {\doibase
  https://doi.org/10.1016/j.physrep.2009.03.003} {\bibfield  {journal}
  {\bibinfo  {journal} {Physics Reports}\ }\textbf {\bibinfo {volume} {476}},\
  \bibinfo {pages} {51 -- 124} (\bibinfo {year} {2009})}\BibitemShut {NoStop}%
\bibitem [{\citenamefont {Toner}\ and\ \citenamefont {Tu}(1998)}]{Toner:1998}%
  \BibitemOpen
  \bibfield  {author} {\bibinfo {author} {\bibfnamefont {John}\ \bibnamefont
  {Toner}}\ and\ \bibinfo {author} {\bibfnamefont {Yuhai}\ \bibnamefont {Tu}},\
  }\bibfield  {title} {\enquote {\bibinfo {title} {Flocks, herds, and schools:
  A quantitative theory of flocking},}\ }\href {\doibase
  10.1103/PhysRevE.58.4828} {\bibfield  {journal} {\bibinfo  {journal} {Phys.
  Rev. E}\ }\textbf {\bibinfo {volume} {58}},\ \bibinfo {pages} {4828--4858}
  (\bibinfo {year} {1998})}\BibitemShut {NoStop}%
\bibitem [{\citenamefont {Gr\'egoire}\ and\ \citenamefont
  {Chat\'e}(2004)}]{gregoire2004onset}%
  \BibitemOpen
  \bibfield  {author} {\bibinfo {author} {\bibfnamefont {Guillaume}\
  \bibnamefont {Gr\'egoire}}\ and\ \bibinfo {author} {\bibfnamefont {Hugues}\
  \bibnamefont {Chat\'e}},\ }\bibfield  {title} {\enquote {\bibinfo {title}
  {Onset of collective and cohesive motion},}\ }\href {\doibase
  10.1103/PhysRevLett.92.025702} {\bibfield  {journal} {\bibinfo  {journal}
  {Phys. Rev. Lett.}\ }\textbf {\bibinfo {volume} {92}},\ \bibinfo {pages}
  {025702} (\bibinfo {year} {2004})}\BibitemShut {NoStop}%
\bibitem [{\citenamefont {Howse}\ \emph {et~al.}(2007)\citenamefont {Howse},
  \citenamefont {Jones}, \citenamefont {Ryan}, \citenamefont {Gough},
  \citenamefont {Vafabakhsh},\ and\ \citenamefont {Golestanian}}]{Howse:2007}%
  \BibitemOpen
  \bibfield  {author} {\bibinfo {author} {\bibfnamefont {Jonathan~R.}\
  \bibnamefont {Howse}}, \bibinfo {author} {\bibfnamefont {Richard A.~L.}\
  \bibnamefont {Jones}}, \bibinfo {author} {\bibfnamefont {Anthony~J.}\
  \bibnamefont {Ryan}}, \bibinfo {author} {\bibfnamefont {Tim}\ \bibnamefont
  {Gough}}, \bibinfo {author} {\bibfnamefont {Reza}\ \bibnamefont
  {Vafabakhsh}}, \ and\ \bibinfo {author} {\bibfnamefont {Ramin}\ \bibnamefont
  {Golestanian}},\ }\bibfield  {title} {\enquote {\bibinfo {title} {Self-motile
  colloidal particles: From directed propulsion to random walk},}\ }\href
  {\doibase 10.1103/PhysRevLett.99.048102} {\bibfield  {journal} {\bibinfo
  {journal} {Phys. Rev. Lett.}\ }\textbf {\bibinfo {volume} {99}},\ \bibinfo
  {pages} {048102} (\bibinfo {year} {2007})}\BibitemShut {NoStop}%
\bibitem [{\citenamefont {Solon}\ \emph {et~al.}(2015)\citenamefont {Solon},
  \citenamefont {Fily}, \citenamefont {Baskaran}, \citenamefont {Cates},
  \citenamefont {Kafri}, \citenamefont {Kardar},\ and\ \citenamefont
  {Tailleur}}]{Solon:2015}%
  \BibitemOpen
  \bibfield  {author} {\bibinfo {author} {\bibfnamefont {A.~P.}\ \bibnamefont
  {Solon}}, \bibinfo {author} {\bibfnamefont {Y.}~\bibnamefont {Fily}},
  \bibinfo {author} {\bibfnamefont {A.}~\bibnamefont {Baskaran}}, \bibinfo
  {author} {\bibfnamefont {M.~E.}\ \bibnamefont {Cates}}, \bibinfo {author}
  {\bibfnamefont {Y.}~\bibnamefont {Kafri}}, \bibinfo {author} {\bibfnamefont
  {M.}~\bibnamefont {Kardar}}, \ and\ \bibinfo {author} {\bibfnamefont
  {J.}~\bibnamefont {Tailleur}},\ }\bibfield  {title} {\enquote {\bibinfo
  {title} {Pressure is not a state function for generic active fluids},}\
  }\href {https://doi.org/10.1038/nphys3377} {\bibfield  {journal} {\bibinfo
  {journal} {Nature Physics}\ }\textbf {\bibinfo {volume} {11}},\ \bibinfo
  {pages} {673} (\bibinfo {year} {2015})}\BibitemShut {NoStop}%
\bibitem [{\citenamefont {Grosberg}\ and\ \citenamefont
  {Joanny}(2015)}]{Grosberg:2015}%
  \BibitemOpen
  \bibfield  {author} {\bibinfo {author} {\bibfnamefont {A.~Y.}\ \bibnamefont
  {Grosberg}}\ and\ \bibinfo {author} {\bibfnamefont {J.-F.}\ \bibnamefont
  {Joanny}},\ }\bibfield  {title} {\enquote {\bibinfo {title} {Nonequilibrium
  statistical mechanics of mixtures of particles in contact with different
  thermostats},}\ }\href {\doibase 10.1103/PhysRevE.92.032118} {\bibfield
  {journal} {\bibinfo  {journal} {Phys. Rev. E}\ }\textbf {\bibinfo {volume}
  {92}},\ \bibinfo {pages} {032118} (\bibinfo {year} {2015})}\BibitemShut
  {NoStop}%
\bibitem [{\citenamefont {Golestanian}\ and\ \citenamefont
  {Ajdari}(2002)}]{Golestanian:2002}%
  \BibitemOpen
  \bibfield  {author} {\bibinfo {author} {\bibfnamefont {R.}~\bibnamefont
  {Golestanian}}\ and\ \bibinfo {author} {\bibfnamefont {A.}~\bibnamefont
  {Ajdari}},\ }\bibfield  {title} {\enquote {\bibinfo {title} {Tracer
  diffusivity in a time- or space-dependent temperature field},}\ }\href
  {http://stacks.iop.org/0295-5075/59/i=6/a=800} {\bibfield  {journal}
  {\bibinfo  {journal} {EPL (Europhysics Letters)}\ }\textbf {\bibinfo {volume}
  {59}},\ \bibinfo {pages} {800} (\bibinfo {year} {2002})}\BibitemShut
  {NoStop}%
\bibitem [{\citenamefont {London}(1938)}]{London:1938}%
  \BibitemOpen
  \bibfield  {author} {\bibinfo {author} {\bibfnamefont {F.}~\bibnamefont
  {London}},\ }\bibfield  {title} {\enquote {\bibinfo {title} {On the
  bose-einstein condensation},}\ }\href {\doibase 10.1103/PhysRev.54.947}
  {\bibfield  {journal} {\bibinfo  {journal} {Phys. Rev.}\ }\textbf {\bibinfo
  {volume} {54}},\ \bibinfo {pages} {947--954} (\bibinfo {year}
  {1938})}\BibitemShut {NoStop}%
\bibitem [{\citenamefont {Kardar}(2007)}]{kardar2007statistical}%
  \BibitemOpen
  \bibfield  {author} {\bibinfo {author} {\bibfnamefont {M.}~\bibnamefont
  {Kardar}},\ }\href {https://books.google.de/books?id=1g8PtAEACAAJ} {\emph
  {\bibinfo {title} {Statistical Physics of Particles}}}\ (\bibinfo
  {publisher} {Cambridge University Press},\ \bibinfo {year}
  {2007})\BibitemShut {NoStop}%
\end{thebibliography}%

\end{document}